 \newcommand{\be}[0]{\begin{equation}}
 \newcommand{\ee}[0]{\end{equation}}
 \newcommand{\ba}[0]{\begin{eqnarray}}
 \newcommand{\ea}[0]{\end{eqnarray}}
\documentstyle[12pt, epsfig]{article}
\begin{document}
\setlength{\topmargin}{-1.5cm}
\vspace*{1.5cm}
\begin{center}

{\LARGE  SOLUTIONS TO THE SCHWINGER-DYSON\\ \vspace{-0.2cm} EQUATION FOR THE
GLUON AND \\ THEIR IMPLICATIONS FOR CONFINEMENT} \\
\vspace{0.8cm}
{\large Kirsten B\"uttner} \\
\vspace{0.5cm}
{\it Centre for Particle Theory, University of Durham,
 Durham DH1 3LE, U.K.} \\

\vspace{1.0cm}

\large {\bf ABSTRACT}

\end{center}

Using the Schwinger-Dyson equations the possible infrared behaviour of the
gluon propagator is studied. Previous work performed in axial gauges
is reviewed, the approximations needed detailed and the difficulties of their
justification discussed. We then turn to the Landau gauge and investigate
the possibility of a gluon propagator less singular than $1/p^{2}$ when
$p^{2} \rightarrow 0$. We find that this infrared softened
behaviour of the gluon propagator is inconsistent; only an
infrared enhanced gluon, as singular as $1/p^{4}$ when
$p^{2} \rightarrow 0$ is consistent with the truncated Schwinger-Dyson
equation. The implications for confinement and for the modelling of the Pomeron
are discussed.

\vspace{2.0cm}

\setlength{\baselineskip}{0.75cm}

\section{Schwinger-Dyson equation approach to QCD}

The complex of Schwinger-Dyson equations (SDE) provides a
non-perturbative framework in which we can
study the infrared properties of QCD and confinement in particular.
(For a comprehensive review see Roberts and Williams~\cite{SDE}.)
Here we describe the infrared behaviour of possible self-consistent solutions
of the truncated SDE for the gluon propagator.

The SDE are the field equations of a quantum field
theory, inter-relating the $n$-point Green's functions. They contain all the
information of the theory but are impossible to solve exactly
since they are an infinite tower of coupled, non-linear integral equations.
In QCD the SDE for the gluon, shown diagrammatically in Fig.~1, gives a
relation for the propagator in terms of the full 3- and 4-point vertex
functions, the quark and ghost propagators and their couplings.
\begin{figure}[thbp]
\begin{center}
\mbox{\epsfig{file=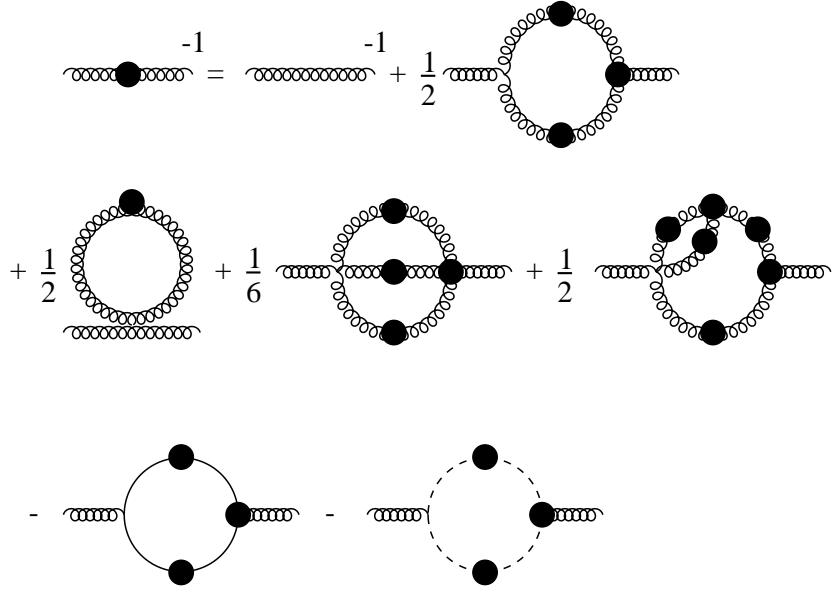,width=12cm}}
\end{center}
\caption[qcd1]{The Schwinger-Dyson equation for the gluon propagator\\
Here the broken line represents the ghost propagator}
\end{figure}
Here we consider a world without quarks, i.e. we neglect the quark
loop in the SDE. This is reasonable since we expect the non-Abelian
nature of QCD to be responsible for confinement.

The infrared behaviour of the gluon propagator has important implications for
quark confinement. A gluon propagator which is as singular
as $1/p^{4}$ when $p^{2} \rightarrow 0$ yields a linearly rising interquark
potential.
Furthermore, West \cite{West} proved, that if, in any gauge,
$\Delta_{\mu\nu}$ is as singular as $1/p^{4}$ then the Wilson operator
satisfies an area law, often regarded as a signal for confinement.
However, another sufficient condition for confinement is the absence of a pole
in the propagator at timelike momenta \cite{RWK}.
So, a gluon propagator which is less singular
than $1/p^{2}$ when $p^{2} \rightarrow 0$ describes a confined  particle.
Such a behaviour was first assumed by
Landshoff and Nachtmann \cite{LN} in their model of the Pomeron.

Extensive work has been done, investigating the infrared behaviour of the
gluon propagator. The different solutions obtained from SDE studies are
displayed diagrammatically in Fig.~2.
\begin{figure}[thbp]
\begin{center}
\mbox{\epsfig{file=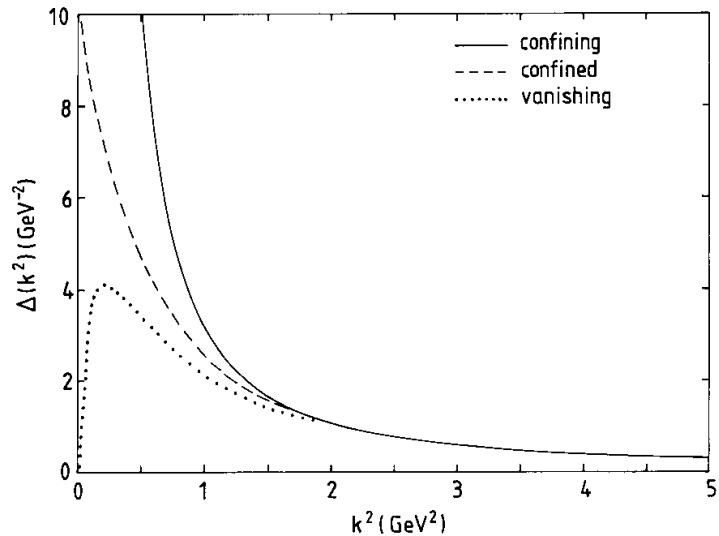,angle=90,width=10cm}}
\end{center}
\caption[qcd2]{Possible behaviour of the gluon propagator $\Delta(p^2)$}
\end{figure}
A number of lattice studies have also been performed, but the infrared
behaviour
of propagators are seriously affected by the lattice's finite size.

The infrared vanishing propagator has been proposed by Stingl et al.
\cite{Stingl} using a different method to solve the approximated set
of SDEs than the one usually employed. However a study of the quark-SDE using
this gluon propagator, \cite{HRW} and \cite{AB}, has shown that
the infrared vanishing gluon propagator does not support dynamical chiral
symmetry breaking and confinement.
Thus the full gluon propagator in QCD cannot have
this behaviour.

We concentrate therefore on the other two solutions: the infrared
enhanced, {\it confining} solution and the {\it confined} solution, which has a
singularity softer than a pole. A {\it confined} gluon has only been claimed
to exist in the axial gauge \cite{CR}, whereas a solution as
singular as $1/p^{4}$ has been shown to exist in both axial and Landau gauges
\cite{BBZ} - \cite{BP}.

\section{Axial gauge studies}

Studies of the axial gauge Schwinger-Dyson equation have the
advantage that ghost fields are absent. However the gluon propagator is
more complicated than in covariant gauges:
\be
 \Delta_{\mu\nu}(p^{2},\gamma) = -\frac{i}{p^{2}}\left[F(p^{2},\gamma)
  M_{\mu\nu} + H(p^{2},\gamma) N_{\mu\nu}\right] \quad ,
\ee
 with the tensors given by:
\[M_{\mu\nu} = g_{\mu\nu} - \frac{p_{\mu}n_{\nu}+p_{\nu}n_{\mu}}
{n\cdot p} + n^{2}\frac{p_{\mu}p_{\nu}}{(n\cdot p)^{2}} \quad , \ \ \mbox{and}
\ \ \ N_{\mu\nu} = g_{\mu\nu} - \frac{n_{\mu}n_{\nu}}{n^{2}} \quad ,\]
and $\gamma=(n\cdot p)^{2}/(n^{2}p^{2})$ is the axial gauge parameter.
However in all previous studies it is assumed that the full propagator has the
same tensor structure as the free one, by setting one of the two axial gauge
gluon renormalisation functions, $H(p^2,\gamma)$, to zero.
As a consequence all 4-gluon terms can be projected out of the
SDE and approximating the 3-gluon vertex by its longitudinal part, which is
determined by the Slavnov-Taylor identity
one finds a closed integral equation for the inverse of the gluon
propagator.

The resulting approximate gluon-SDE has been studied numerically by Baker,
Ball and Zachariasen (BBZ) \cite{BBZ} and an infrared enhanced gluon
propagator, as singular as $1/p^4$ was found to be a consistent solution.
However it has been argued \cite{West2} that in axial gauges, in which only
positive norm-states occur, a behaviour more singular than $1/p^2$ is not
possible and consequently the neglected axial gauge renormalisation function
$H$ must cancel any $1/p^4$ singularity in the infrared.
More recently, Cudell and Ross \cite{CR} have shown that an alternative
solution
to Schoenmaker's approximation \cite{Schoen} to BBZ's equation exists with an
infrared softened gluon propagator. However this solution is found only with an
incorrect sign in Schoenmaker's equation. Correcting this error, the axial
gauge
SDE does not allow an infrared softened solution for the propagator
\cite{physrev}.

Because of the difficulty in justifying the neglect of one of the key gluon
renormalisation functions in axial gauges, we turn our attention to covariant
gauges and the Landau gauge in particular.

\section{Landau gauge studies}

The advantage of Landau gauge studies is the much simpler structure of
the gluon propagator, which is defined by~:
\be
\Delta_{\mu\nu} = - i\ \frac{G(p^2)}{p^2}\left(g_{\mu\nu} - \frac{p_{\mu}
p_{\nu}}{p^2}\right)\qquad .
\ee
\noindent However in any covariant gauge ghosts are necessary to ensure the
transversality
of the gluon --- nothing comes for free! The gluon-SDE is approximated
by neglecting all 4-gluon terms, which can be regarded as the first
step in a truncation of the SDE, as discussed by Mandelstam \cite{Mandel}, and
including ghosts only perturbatively. The latter leads to an approximation of
the Slavnov-Taylor identity, which is now treated as in the Abelian case.
Again we replace the 3-gluon vertex by its longitudinal part determined by
this Slavnov-Taylor identity.
The resulting truncated Schwinger-Dyson equation is:
\ba
\lefteqn{\frac{1}{G(p^2)} =
1 + \frac{g^{2}C_{A}}{96\pi^4}\frac{1}{p^2}\int d^4k \left[ G(q^2)
A(k^2,p^2) + \frac{G(k^2)G(q^2)}{G(p^2)}\ B(k^2,p^2)\right.} \nonumber \\
& & \qquad \quad + \left.\frac{G(k^2)-G(p^2)}
{k^2-p^2}\frac{G(q^2)}{G(p^2)}\ C(k^2,p^2) + \frac{G(q^2)-G(k^2)}
{q^2-k^2} \ D(k^2,p^2) \right] ,
\ea
\noindent with $q=p-k$ and where $A$, $B$, $C$ and $D$ are functions of $p^2$
and $k^2$ as given in \cite{BP}.

Brown and Pennington \cite{BP} studied this equation numerically and
found an infrared singular, {\it confining} gluon to be a consistent solution.
To allow an analytic study of the Brown-Pennington equation, we
approximate $G(q^2)$ by $G(p^2+k^2)$, as first proposed by Schoenmaker
\cite{Schoen}. This should be exact in the infrared
limit. Now the angular integrals can be performed analytically giving the
following simpler equation:
\ba
\lefteqn{\frac{1}{G(p^2)} = 1 + } \nonumber \\
& &\frac{g^{2} C_{A}}{48\pi^2}\frac{1}{p^2}\left\{
\int_{0}^{p^2} dk^2 \left[ G_{1} \left( -1 - 10\frac{k^2}{p^2}
+6\frac{k^4}{p^4} + \frac{k^2}{p^2-k^2} \left( \frac{75}{4} -
\frac{39}{4}\frac{k^2}{p^2} + 4\frac{k^4}{p^4} - 5\frac{p^2}{k^2}\right)
\right) \right.\right. \nonumber \\
& & \left. \qquad \qquad + G_2 \left(-\frac{21}{4}\frac{k^2}{p^2} + 7
\frac{k^4}{p^4} - 3 \frac{k^6}{p^6}\right) +
G_3 \left( \frac{k^2}{p^2-k^2}\left( -\frac{27}{8} - \frac{11}{4}
\frac{k^2}{p^2} - \frac{15}{8}\frac{p^2}{k^2} \right)\right)\right]
\nonumber \\
& & \qquad \qquad +\int_{p^2}^{\infty} dk^2 \left[ G_{1} \left( \frac{p^2}{k^2}
 - 6 + \frac{p^2}{p^2-k^2}\left( \frac{29}{4} + \frac{3}{4}\frac{p^2}{k^2}
\right)\right) \,+ \right.\nonumber \\
& & \qquad \qquad \left.\left. + G_2 \left( - \frac{3}{2} + \frac{1}{4}
\frac{p^2}{k^2} \right) +
G_3 \left( \frac{p^2}{p^2-k^2}\left( \frac{3}{4} -
\frac{67}{8}\frac{p^2}{k^2} - \frac{3}{8}\frac{p^4}{k^4} \right)\right)
\right] \right\} \, ,
\ea
where \[G_{1} = G(p^2+k^2) \;, \quad
G_{2} =  G(p^2+k^2)- G(k^2) \quad \mbox{and} \quad
G_{3} = \frac {G(k^2)G(p^2+k^2)}{G(p^2)} \; . \]

This equation has a quadratic ultraviolet divergence, which
would give a mass to the gluon. Such terms are subtracted to ensure
the masslessness condition
\be \lim_{p^2 \rightarrow 0} {1\over \Delta(p^2)} = 0 \ \ \mbox{, i.e.}\ \ \
\frac{p^2}{G(p^2)} = 0 \ \ \mbox{for}\ \ p^2 \rightarrow 0 \qquad ,\ee
is satisfied. This property can be derived generally from the Slavnov-Taylor
identity and always has to hold. To determine the possible self-consistent
behaviour of the gluon renormalisation function, $G(p^2)$ is expanded in a
series in powers of $p^2/\mu^2$ for $p^2 < \mu^2$ (including possible negative
powers). Here we describe only the lowest powers for illustration.
$\mu^2$ is the mass scale above
which we assume perturbation theory applies and  we demand that for  $p^2 >
\mu^2$ the solution of the integral equation matches the perturbative
result, i.e. we have $G(p^2) = 1$ modulo logarithms.

To ensure that making Schoenmaker's approximation \cite{Schoen} does not
qualitatively alter the behaviour of Eq.~(1) we first reproduce the infrared
enhanced {\it confining} gluon found by Brown and Pennington.
Taking (cf. \cite{BP})
\[G_{in}(k^2) = A \left(\frac{\mu^2}{k^2}\right)_+ \]
as an input function, where the $+$ serves as an infrared regulator, we find:
\[ \frac{1}{G_{out}(p^2)} = 1 + \ \frac{\mbox{const}}{p^2} \qquad .\]

\noindent This violates the masslessness condition of Eq.~(5) and
has to be mass renormalised. Then, if terms in $G(p^2)$, of higher order in
$p^2$, give a contribution to the right hand side of the equation
cancelling the explicit factor of 1, we have the possibility of finding
self-consistency. Consequently, we take
\be
G_{in}(p^2) = \left\{ \begin{array}{ll}
A \left({\mu^2}/{p^2} \right)_+ + \left({p^2}/{\mu^2} \right)
& \mbox{if $p^2 < \mu^2 $} \\ \\
1 & \mbox{if $p^2 > \mu^2 $}
\end{array}
\right. \qquad ,\ee
\noindent and find after mass renormalisation:
\be
\frac{1}{G_{out}(p^2)} = 1 + \frac{g^{2}C_{A}}{48\pi^2} \left[
-\frac{479}{24}\frac{p^2}{\mu^2} +
\frac{13}{8}\frac{p^2}{\mu^2}\ln\left(\frac{\mu^2}{p^2}\right)
-\frac{81}{4} -
\frac{25}{4}\ln\left(\frac{\Lambda^2}{\mu^2}\right) \right] \qquad .\ee

\noindent Hence an infrared enhanced {\it confining} gluon propagator is a
self-consistent solution to the SDE.
This is the result found by Brown and Pennington \cite{BP}. Higher powers of
$p^2$ in Eq.~(6) do not affect the answer.

To check whether Eq.~(2) allows an infrared softened gluon propagator, i.e. an
infrared vanishing renormalisation function, we take (cf.
\cite{CR})
\be
G_{in}(p^2) = \left\{ \begin{array}{ll}
\left({p^2}/{\mu^2}\right)^{1-c} & \mbox{if $p^2< \mu^2$} \\ \\
1 & \mbox{if $p^2> \mu^2$}
\end{array}
\right. \ee
\noindent as a trial input function and substitute it into the right hand side
of the integral equation, Eq.~(2).
Performing the $k^2$-integration, we obtain, after mass renormalisation:
\ba
\frac{1}{G_{out}(p^2)} &=& 1 + \frac{g^2C_{A}}{48\pi^2} \left[
\ D_{1} + \ D_{2} \left( \frac{\mu^2}{p^2}\right)^{1-c}
+ \ D_{3} \left(\frac{p^2}{\mu^2} \right)^{1-c} +\  D_{4}\left(
\frac{p^2}{\mu^2}\right)^{c} + ...\right]  \ ,
\ea
\noindent where $G_{1}, G_{2}$ and $G_{3}$ have been expanded
for small $p^2$ and only the first few terms have been collected in this
equation so that
\be
D_{2} = -\left( \frac{3}{4(2-2c)} +\frac{3}{4}\ln\left(
\frac{\Lambda^2}{\mu^2} \right) \right)\nonumber \qquad .
\ee
\noindent The other $D$'s can be found in \cite{physrev}.
Thus the dominant infrared behaviour is:
\be
\frac{1}{G_{out}(p^2)} \rightarrow - \left(\frac{\mu^2}{p^2}
\right)^{1-c}
\ee
and self-consistency is spoiled by a negative sign, just as in axial gauges
\cite{physrev}. Note that higher order terms in $p^2$ in the input function
have no qualitative effect.
We thus see that an infrared softened {\it confined} gluon is not possible.

\section{Consequences for modelling the Pomeron}

How does this infrared behaviour of the gluon affect the Pomeron of Landshoff
and Nachtmann \cite{LN}? Their belief in an infrared
softened, rather than enhanced, gluon rests on their model requirement that the
integral \[ \int^{\infty}_{0} dp^2 \Delta(p^2)^2 \] should be finite. However,
we do not believe that whether this integral is finite
or not is relevant to the finiteness of total cross-sections \cite{pom}.
The Landshoff-Nachtmann picture imagines that the two dressed gluons
modeling their Pomeron couple to single quarks with the other quarks in each
initial state hadron being spectators. In this way the forward hadronic
scattering amplitude is viewed as quark-quark scattering.
The total cross-section is then just the
imaginary part of this forward elastic quark scattering amplitude, by the
optical theorem. However, an imaginary part is only generated if the quarks can
be on mass-shell and have poles in their propagators which is in conflict with
confinement. Hadronic amplitudes are not just the result of free quark
interactions, perturbative QCD is only valid for hard short distance processes.
In soft, non-perturbative physics, the bound state nature of hadrons has to be
taken into account.

\section{Summary}

We have studied the Schwinger-Dyson equation of the gluon propagator in the
Landau gauge to determine analytically the possible infrared solutions for the
gluon renormalisation function $G(p^2)$. We find only an infrared enhanced
gluon, as singular as $1/p^4$ for $p^2 \rightarrow 0$ is consistent with the
truncated SDE. This behaviour of the gluon is not at variance with the Pomeron
and is in accord with quark confinement.
Furthermore it leads to a good phenomenology of hadron observables
\cite{SDE}, \cite{Craig}.

\section{ Acknowledgements}

This work was performed in collaboration with Michael Pennington. I wish to
thank the University of Durham for a research studentship and the ELFE Project
for providing me with the funds to attend this School.

\pagebreak

\end{document}